\pdfoutput=1

\documentclass[conference]{IEEEtran}
\IEEEoverridecommandlockouts
\usepackage{cite}
\usepackage{amsmath,amssymb,amsfonts}
\usepackage[bookmarks,colorlinks,citecolor=green]{hyperref}
\usepackage{algorithmic}
\usepackage{acronym}
\usepackage[tableposition=top]{caption}
\usepackage{graphicx}
\usepackage{textcomp}
\usepackage{xcolor}
\usepackage[pscoord]{eso-pic}

\newcommand{\myVec}{\ignorespaces}

\begin{document}

\acrodef{fpn}[FPN]{fixed-pattern noise}
\acrodef{lrr}[LRR]{low-rank regularization}
\acrodef{nlm}[NLM]{non-local means}
\acrodef{gf}[GF]{guided-filter}
\acrodef{mhe}[MHE]{midway histogram equalization}
\acrodef{snrwdnn}[SNRWDNN]{stripe noise removal wavelet deep neural network}
\acrodef{cnn}[CNN]{convolutional neural network}
\acrodef{snrcnn}[SNRCNN]{stripe noise removal convolutional neural network}
\acrodef{st-snr}[ST-SNR]{spatiotemporal stripe noise removal}
\acrodef{bigcru}[BiGCRU]{bidirectional gated convolutional recurrent units}
\acrodef{bigru}[BiGRU]{bidirectional gated recurrent units}
\acrodef{lrsid}[LRSID]{low-rank single-image decomposition}
\acrodef{psnr}[PSNR]{peak signal-to-noise ratio}
\acrodef{ssim}[SSIM]{structural similarity index}
\acrodef{dinr}[DINR]{deep-unfolding for iterative noise removal}
\acrodef{rnn}[RNN]{recurrent neural network}

\newcommand{\placetextbox}[3]{
  \setbox0=\hbox{#3}
  \AddToShipoutPictureFG*{
    \put(\LenToUnit{#1\paperwidth},\LenToUnit{#2\paperheight}){\vtop{{\null}\makebox[0pt][c]{#3}}}%
  }%
}%

\title{Deep Unfolding for Iterative Stripe Noise Removal\\
}

\author{\IEEEauthorblockN{{Zeshan Fayyaz\textsuperscript{1}}}
\IEEEauthorblockA{ 
\textit{zeshan.fayyaz@ryerson.ca}}
\and
\IEEEauthorblockN{Daniel Platnick\textsuperscript{1}}
\IEEEauthorblockA{ 
\textit{daniel.platnick@ryerson.ca}}
\and
\IEEEauthorblockN{Hannan Fayyaz\textsuperscript{2}}
\IEEEauthorblockA{ 
\textit{hannanfayyaz@gmail.com}}
\and
\IEEEauthorblockN{Nariman Farsad\textsuperscript{1}}
\IEEEauthorblockA{ 
\textit{nfarsad@ryerson.ca}}

}

\maketitle

\placetextbox{0.5}{0.84}{\normalsize{\textsuperscript{1}Ryerson University, \textsuperscript{2}York University}}

\begin{abstract}
The non-uniform photoelectric response of infrared imaging systems results in fixed-pattern stripe noise being superimposed on infrared images, which severely reduces image quality. As the applications of degraded infrared images are limited, it is crucial to effectively preserve original details. Existing image destriping methods struggle to concurrently remove all stripe noise artifacts, preserve image details and structures, and balance real-time performance. In this paper we propose a novel algorithm for destriping degraded images, which takes advantage of neighbouring column signal correlation to remove independent column stripe noise. This is achieved through an iterative deep unfolding algorithm where the estimated noise of one network iteration is used as input to the next iteration. This progression substantially reduces the search space of possible function approximations, allowing for efficient training on larger datasets. The proposed method allows for a more precise estimation of stripe noise to preserve scene details more accurately. Extensive experimental results demonstrate that the proposed model outperforms existing destriping methods on artificially corrupted images on both quantitative and qualitative assessments.  
\end{abstract}


\section{Introduction}
\label{sec:intro}
Infrared imaging systems are an important tool used across many field domains, including medical imaging, transport navigation, and remote sensing \cite{kuangsingle}. Infrared images are typically corrupted by stripe noise due to the non-uniform sensing of light in the system’s photo-receptive sensors \cite{WANG201658}. This corruption results in significant \ac{fpn} embedded in the image, which decreases the quality of infrared imaging systems. To produce a more accurate image, it is imperative to remove the superimposed vertical stripe noise artifacts and preserve the original structures of the image.

Previous destriping methods can be placed into three categories: optimization-based methods, statistics-based methods, and deep learning-based methods. Optimization-based stripe noise correction methods are contextualized as an ill-posed inverse problem, where several priors are inputted into the regularizer model \cite{sym10050157}. \Ac{lrr} \cite{chang2016remote}, \ac{nlm} \cite{li2016novel} and \ac{gf} \cite{he2012guided} are methods that use prior knowledge of the ground truth to remove stripe noise. Prior-based strategies remove stripe noise indiscriminately, resulting in blurred image artifacts. Standard statistics-based methods include the \ac{mhe} \cite{MHE} approach. This algorithm evenly distributes pixel intensity values throughout the image, typically increasing contrast and image clarity. One drawback to the \ac{mhe} algorithm is that due to its indiscriminate nature, it may increase the contrast of noise artifacts and hinder the image signal quality.

Deep learning-based methods eventually show vast improvement in the performance of stripe-noise removal algorithms. J. Guan et al.’s \cite{guan2019wavelet} \ac{snrwdnn} consists of a \ac{cnn} which predicts the wavelet transform coefficients of an image, and then the inverse transformation is applied to achieve the destriped image. Additionally, J. Guan et al.’s \cite{guan2019_spatiotemp} \ac{st-snr} approach uses \ac{bigcru} to take advantage of the strong dependency between the continuous stripe component along the columns and the rows. These methods are effective at removing low to medium levels of stripe noise but still leave minor stripe artifacts, especially when corrupted with higher levels of noise. 

As the applications of degraded infrared images are limited, it is crucial to remove the column-wise noise while preserving complex details. Previous traditional destriping methods, such as low-rank and sparse matrix decomposition often lead to inaccurate sparse modeling and unstable results. Deep learning-based denoising methods originate from unsupervised low-rank sparse decomposition for feature extraction \cite{song2021unsupervised}. Y. Wan et al. \cite{wan2021accurate} present a deep-learning based destriping approach based on an accurate multi-objective low-rank sparse denoising framework, and the problem is converted into a multi-objective optimization problem. 

Inspired by deep-unfolding techniques \cite{hershey2014_deepUnfold, monga2021_deepUnfold}, to overcome this challenge, in this work, we propose the \ac{dinr} algorithm. In the proposed method, \ac{rnn}s are used to iteratively remove column noise from the image. In particular, during each iteration, the noise over each column is estimated using the current as well as the adjacent columns. The high correlation between adjacent columns in the clean image can be used by the algorithm to better differentiate the noise from the original signal.  The estimated noise at the end of each iteration is then used to progressively clean the image. Thus, our method continuously feeds the output of the network, which is a partially destriped image, as input back into the network. This iterative noise estimation and removal destripes an image until all stripe noise artifacts are removed. In certain instances, DINR outperforms the current state-of-the-art (SOTA) for stripe noise removal by 22.99\% on quantitative assessments, as well as qualitatively more accurately preserves complex scene details and original shadowing in high-intensity noise regions. 

Deep learning approaches attempt to discover model information by optimizing network parameters learned from training. Although highly efficient, deep learning typically suffers the drawbacks of requiring large training sets, lack of interpretability, and overfitting. In general, RNNs generate predictions over many time steps, which can be simplified and further improved by unfolding, or unrolling the algorithm over the input sequence. Unfolded networks inherit prior domain and structure knowledge, rather than learnt through extensive training and are capable of more accurately approximating the target function due to its universal approximation capability \cite{monga2021algorithm}. Deep unrolled networks have previously been deployed for video super-resolution tasks, as explored in the working of B. N. Chiche et al. \cite{chiche2020deep} and results find that unrolled networks allow for flexibility in learning a single model to nonblindly deal with multiple degradations while learning spatial patterns and details.  

The rest of this paper is organized as follows. The problem statement is discussed in Section \ref{sec:problem_statement}. Then, in Section \ref{sec:method}, we present \ac{dinr}. Section \ref{sec:results} presents the evaluation results and comparison to prior work, and the paper ends with concluding remarks in Section \ref{sec:conclusion}.

\section{STRIPE NOISE REMOVAL PROBLEM AND MOTIVATION}
\label{sec:problem_statement}

\begin{figure}[!b]

\begin{minipage}[!b]{1.0\linewidth}
  \centering
  \centerline{\includegraphics[width=8.5cm]{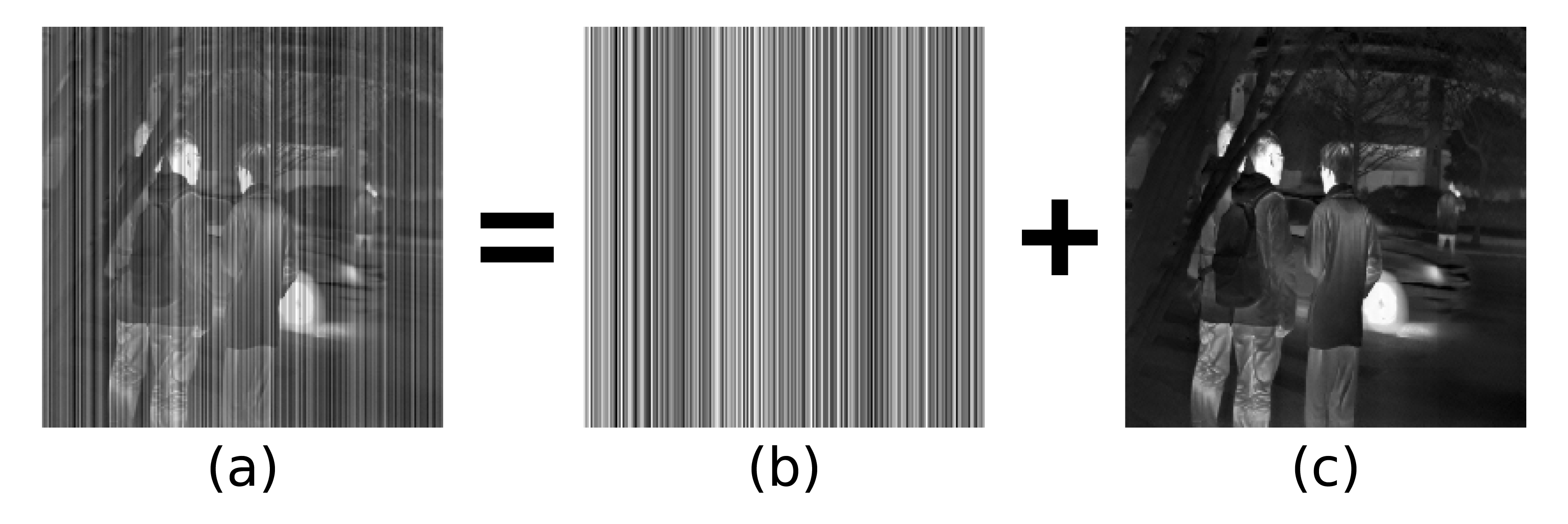}}
\end{minipage}
\caption{An example of a noisy infrared image, separated from it's FPN. (a) degraded image; (b) stripe component; (c) clean image.}
\label{fig:stripeNoise}
\end{figure}


 \begin{figure*}[htb]
     \centering
     \includegraphics[width =\textwidth]{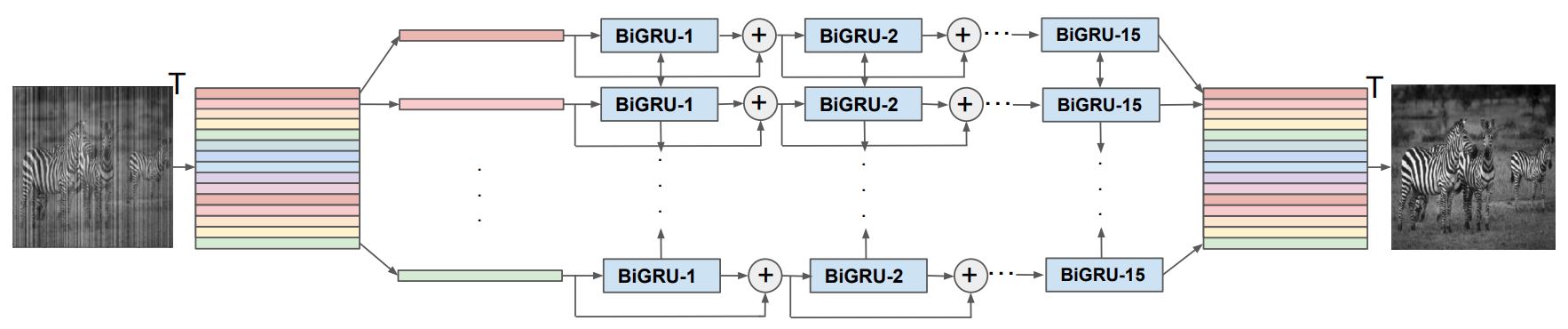}
     \caption{The network architecture of the proposed DINR method. Every pixel column of the noisy image is transposed and input to a first-layer BiGRU cell, where the number of cells is the number of columns in the image. Each iteration in the iterative denoising algorithm is captured by a BiGRU layer.}
     \label{fig:dinrArch}
 \end{figure*}
 

\begin{table*}[]
\centering
\normalsize
\caption{Average PSNR and SSIM results for datasets Urban100, Set12, Infrared100, Linnaeus 5, and BSDS100 for various network architectures.}
\begin{tabular}{c|cc|cc|cc|cc}
      & \multicolumn{2}{c|}{$\beta$ = 0.05} & \multicolumn{2}{c|}{$\beta$ = 0.15} & \multicolumn{2}{c|}{$\beta$ = 0.25} & \multicolumn{2}{c}{Average} \\
Model & PSNR          & SSIM          & PSNR          & SSIM          & PSNR          & SSIM          & PSNR         & SSIM         \\ \hline
14 BiGRU & 35.70         & 0.991         & 32.48         & 0.985         & 31.89         & 0.983         & 33.11        & 0.986        \\
\textbf{15 BiGRU} & \textbf{36.02}         & \textbf{0.991}         & \textbf{32.61}         & \textbf{0.986}         & \textbf{31.17}         & \textbf{0.982}         & \textbf{33.27}        & \textbf{0.986}        \\
16 BiGRU & 35.78         & 0.991         & 32.33         & 0.985         & 31.08         & 0.982         & 33.06        & 0.986        \\
17 BiGRU & 35.10         & 0.990         & 31.79         & 0.983         & 30.44         & 0.978         & 32.44        & 0.984        \\
18 BiGRU & 35.90         & 0.991         & 32.39         & 0.985         & 30.92         & 0.982         & 33.07        & 0.986       
\end{tabular}
\label{tbl:ablation}
\end{table*}

\begin{table*}[]
\small
\caption{Mean PSNR and SSIM results of various classical destriping methods on Set12. \ac{dinr} achieves a 9.73\% increase in PSNR and 1.13\% increase in SSIM, as compared to the previous SOTA.}
\centering
\begin{tabular}{l|llllllllllll}
                    & \multicolumn{2}{c}{$\beta$=0.06}        & \multicolumn{2}{c}{$\beta$=0.1}         & \multicolumn{2}{c}{$\beta$=0.14}        & \multicolumn{2}{c}{$\beta$=0.18}        & \multicolumn{2}{c}{$\beta$=0.22}                             & \multicolumn{2}{c}{Average}     \\
                    & PSNR           & SSIM           & PSNR           & SSIM           & PSNR           & SSIM           & PSNR           & SSIM           & PSNR           & \multicolumn{1}{l|}{SSIM}           & PSNR           & SSIM           \\ \hline
Corrupted           & 24.54          & 0.750          & 20.10           & 0.564           & 17.18           & 0.432           & 15.00           & 0.339          & 13.25         & \multicolumn{1}{l|}{0.271}          & 18.01          & 0.471           \\
GF \cite{he2012guided}                 & 27.36          & 0.803          & 24.91          & 0.704          & 23.05          & 0.629          & 21.68          & 0.587          & 20.71          & \multicolumn{1}{l|}{0.555}          & 23.54          & 0.656          \\
NLM \cite{li2016novel}                & 26.97          & 0.778          & 24.48          & 0.659          & 22.86          & 0.573          & 21.37          & 0.503          & 20.24          & \multicolumn{1}{l|}{0.445}          & 23.18          & 0.592          \\
MHE \cite{MHE}                 & 27.86          & 0.879          & 24.91          & 0.812          & 22.45          & 0.747          & 20.46          & 0.680          & 18.77          & \multicolumn{1}{l|}{0.613}          & 22.89          & 0.746          \\
LRSID \cite{chang2016remote}              & 30.63          & 0.943          & 29.42          & 0.938          & 27.90          & 0.927          & 26.23          & 0.908          & 24.51          & \multicolumn{1}{l|}{0.874}          & 27.74          & 0.918          \\
SNRCNN \cite{kuangsingle}             & 28.13          & 0.944          & 26.44          & 0.929          & 24.79          & 0.906          & 23.28          & 0.878          & 21.93          & \multicolumn{1}{l|}{0.842}          & 24.91          & 0.900          \\
DLSNUC \cite{he2018single}            & 28.46          & 0.950          & 26.56          & 0.940          & 25.01          & 0.918          & 23.45          & 0.898          & 22.13          & \multicolumn{1}{l|}{0.873}          & 25.12          & 0.916          \\
ICSRN \cite{xiao2018removing}              & 28.73          & 0.958          & 26.98          & 0.947          & 25.26          & 0.931          & 23.72          & 0.911          & 22.36          & \multicolumn{1}{l|}{0.887}          & 25.41          & 0.927          \\
SNRWDNN \cite{guan2019wavelet}            & 33.18          & 0.988          & 30.07          & 0.982          & 28.01          & 0.976          & 26.43          & 0.970          & 25.12          & \multicolumn{1}{l|}{0.964}          & 28.56          & 0.976          \\ \hline
\textbf{Our Method} & \textbf{33.57} & \textbf{0.990} & \textbf{31.89} & \textbf{0.988} & \textbf{30.95} & \textbf{0.988} & \textbf{30.36} & \textbf{0.986} & \textbf{29.95} & \multicolumn{1}{l|}{\textbf{0.985}} & \textbf{31.34} & \textbf{0.987}
\end{tabular}
\label{tbl:results12Data}
\end{table*}

To model the stripe noise in infrared imaging, we follow the models proposed in prior work~\cite{guan2019wavelet, guan2019_spatiotemp}. Let $\myVec{X}$, $\myVec{S}$, and $\myVec{Y}$, respectively, denote the $n \times m$ matrices of the original clean image, the stripe noise, and the degraded image. Then the noise added to $i$-th column is given by: 

\begin{equation}
\label{eq:noise}
    \myVec{y}_i = \myVec{x}_i + \myVec{s}_i,    
\end{equation}
where $\myVec{y}_i$, $\myVec{x}_i$, and $\myVec{s}_i$ are the $i$-th column of $\myVec{Y}$, $\myVec{X}$, and $\myVec{S}$, respectively, and the elements of $\myVec{s}_i$ are equal in value (i.e., $s_i^{(1)}=s_i^{(2)} \cdots = s_i^{(n)}$ and distributed as 
\begin{equation}
    s_i \sim N(0, \sigma^2).    
\end{equation}
While the noise variance remains the same across the columns of the same image, the variance can change from image to image. Specifically, for the stripe matrix $\myVec{S}$, the standard deviation is distributed as
\begin{equation}
\label{eq:betaDist}
    \sigma \sim U(0, \beta),
\end{equation}
where $U$ is the uniform distribution, and $\beta$ controls the noise power. Fig.~\ref{fig:stripeNoise} shows a clean sample image, stripe noise, and the noisy image. 

 The goal of denoising is to estimate the clean image $\hat{\myVec{X}}$ from the noisy observation $\myVec{Y}$. Specifically, inspired by deep-unfolding techniques \cite{hershey2014_deepUnfold, monga2021_deepUnfold}, our proposed algorithm aims to iteratively estimate the residual noise and progressively clean the image.

\section{Deep Unfolding for Iterative Noise Removal}

\label{sec:method}
In this section, we will introduce the overall network architecture and outline our approach to stripe noise removal. 

Given a noisy image $\myVec{Y}$, our proposed algorithm \ac{dinr} aims to estimate the noise and the clean image iteratively. Specifically, let $\hat{\myVec{X}}^{(k)}$ be the estimated clean image at the end of the $k$-th iteration with $\hat{\myVec{X}}^{(0)} = \myVec{Y}$, and let $\hat{\myVec{S}}^{(k)}$ be the estimated noise at the end of the $k$-th iteration. Instead of our algorithm estimating the clean image directly during each iteration, it estimates the noise. Therefore, the estimated noise after the $k$-th iteration is given by 
\begin{align}
\label{eq:resNoise}
  \hat{\myVec{S}}^{(k)} = \hat{\myVec{X}}^{(0)} - \hat{\myVec{X}}^{(k)}.  
\end{align}
During the iteration, the noise is re-estimated from the previous iteration using a function $\hat{\myVec{S}}^{(k)} = f_k (\hat{\myVec{X}}^{(k-1)})$. Therefore, the output of the $k$-th iteration is given by:
\begin{align}
\label{eq:iterStep}
    \hat{\myVec{X}}^{(k)} = \hat{\myVec{X}}^{(0)} - f_k (\hat{\myVec{X}}^{(k-1)}).
\end{align}

To design the function $f_k$ that estimates the residual noise from the previous step, we use the following facts about stripe noise. 1) The same noise is added to every pixel in a given column of the image. 2) To distinguish the noise from ground truth pixel values of the clean image for the $i$-th column, pixel values from adjacent columns can be exploited as they will be highly correlated with the $i$-th column. 

Using these two facts, we use \ac{rnn}s, specifically \ac{bigru}, to represent the function $f_k$. That is $f_k$ is represented by the $k$-th layer of a multi-layer \ac{bigru}, where the \ac{bigru} inputs are $m$ vectors of length $n$. Therefore, the function $f_k$ in \eqref{eq:iterStep}, for the $i$-th column is given by:
\begin{align}
\label{eq:BiGRU}
    f_k (\hat{\myVec{x}}_i^{(k-1)}) = \text{BiGRU}_k \left( \hat{\myVec{x}}_i^{(k-1)}, g_k(\hat{\myVec{x}}_{0:i-1}^{(k-1)}), h_k(\hat{\myVec{x}}_{i+1:m}^{(k-1)}) \right),
\end{align}
where $g_k(.)$ and $h_k(.)$ are the GRU states from the forward and backward GRUs, respectively. These state vectors summarize relevant information from the columns before and the columns after the $i$-th column. Using this technique, the network can denoise an image column-wise using spatial information from neighbouring columns. A GRU is a modified type of \ac{rnn}. As opposed to RNNs, GRUs merge the input gate and the forget gate into a single update gate and merge the cell state into the hidden state \cite{YinBiGRU}. A BiGRU extensively gathers redundant information from past and future inputs to better estimate the stripe component. The bidirectional strategy allows us to compare columns with both of its neighbours, strengthening its long-time correlation and allowing for learning of temporal and spatial contextual information simultaneously. 


Unlike the BiGCRU proposed in J. Guan et. al  \cite{guan2019_spatiotemp}, a BiGRU better preserves complex scene details by not over-smoothing an image with a convolutional layer. A comprehensive ablation study performed with BiGRU layers stacked with convolutional layers can be found in our source code repository.  Our overall algorithm is shown in Fig.~\ref{fig:dinrArch}. 
Assuming we unfold the algorithm for $T$ iterations, the deep-unfolded layers can then be trained end-to-end using a mean squared error loss between $\hat{\myVec{X}}^{(T)}$ and $\myVec{X}$.

\begin{table}[!htbp]
\small
\caption{Mean PSNR and SSIM results of various test datasets against SNRWDNN. \ac{dinr} achieves an average increase of 22.99\% and 3.79\% for PSNR and SSIM, respectively.}
\centering
\begin{tabular}{llcccc}
                                 & \multicolumn{1}{l|}{}                     & \multicolumn{2}{c}{SNRWDNN} & \multicolumn{2}{c}{Our Method}    \\
\multicolumn{1}{l|}{Dataset}     & \multicolumn{1}{c|}{$\beta$} & PSNR          & SSIM         & PSNR             & SSIM           \\ \hline
\multicolumn{1}{l|}{}            & \multicolumn{1}{l|}{0.05}                 & 32.32         & 0.981        & \textbf{38.12}   & \textbf{0.995} \\
\multicolumn{1}{l|}{BSDS100}     & \multicolumn{1}{l|}{0.15}                 & 26.36         & 0.957        & \textbf{36.15}   & \textbf{0.993} \\
\multicolumn{1}{l|}{}            & \multicolumn{1}{l|}{0.25}                 & 22.73         & 0.922        & \textbf{34.44}   & \textbf{0.991} \\ \hline
\multicolumn{1}{l|}{}            & \multicolumn{1}{l|}{0.05}                 & 33.53         & 0.980        & \textbf{38.42}   & \textbf{0.992} \\
\multicolumn{1}{l|}{INFRARED100} & \multicolumn{1}{l|}{0.15}                 & 27.04         & 0.950        & \textbf{34.67}   & \textbf{0.986} \\
\multicolumn{1}{l|}{}            & \multicolumn{1}{l|}{0.25}                 & 23.03         & 0.899        & \textbf{32.79}            & \textbf{0.982}          \\ \hline
\multicolumn{1}{l|}{}            & \multicolumn{1}{l|}{0.05}                 & 32.36         & 0.983        & \textbf{34.29}   & \textbf{0.991} \\
\multicolumn{1}{l|}{Set12}       & \multicolumn{1}{l|}{0.15}                 & 26.09         & 0.959        & \textbf{30.77}   & \textbf{0.987} \\
\multicolumn{1}{l|}{}            & \multicolumn{1}{l|}{0.25}                 & 22.41         & 0.922        & \textbf{29.71}   & \textbf{0.982} \\ \hline
\multicolumn{1}{l|}{}            & \multicolumn{1}{l|}{0.05}                 & 32.66         & 0.982        & \textbf{35.70} & \textbf{0.990} \\
\multicolumn{1}{l|}{Linnaeus 5}  & \multicolumn{1}{l|}{0.15}                 & 26.59         & 0.958        & \textbf{31.73}   & \textbf{0.983} \\
\multicolumn{1}{l|}{}            & \multicolumn{1}{l|}{0.25}                 & 22.90         & 0.921        & \textbf{30.39}   & \textbf{0.979} \\ \hline
\multicolumn{1}{l|}{}            & \multicolumn{1}{l|}{0.05}                 & 30.63         & 0.978        & \textbf{33.58}   & \textbf{0.988} \\
\multicolumn{1}{l|}{Urban100}    & \multicolumn{1}{l|}{0.15}                 & 25.16         & 0.948        & \textbf{29.75}   & \textbf{0.980} \\
\multicolumn{1}{l|}{}            & \multicolumn{1}{l|}{0.25}                 & 22.00         & 0.915        & \textbf{28.52}   & \textbf{0.975} \\ \hline
Average                          &                                           & 27.05         & 0.950        & \textbf{33.27}   & \textbf{0.986}
\end{tabular}
\label{tbl:resultsAllDataset}
\end{table}
 
  \begin{figure*}[htb]
     \centering
     \includegraphics[width =\textwidth]{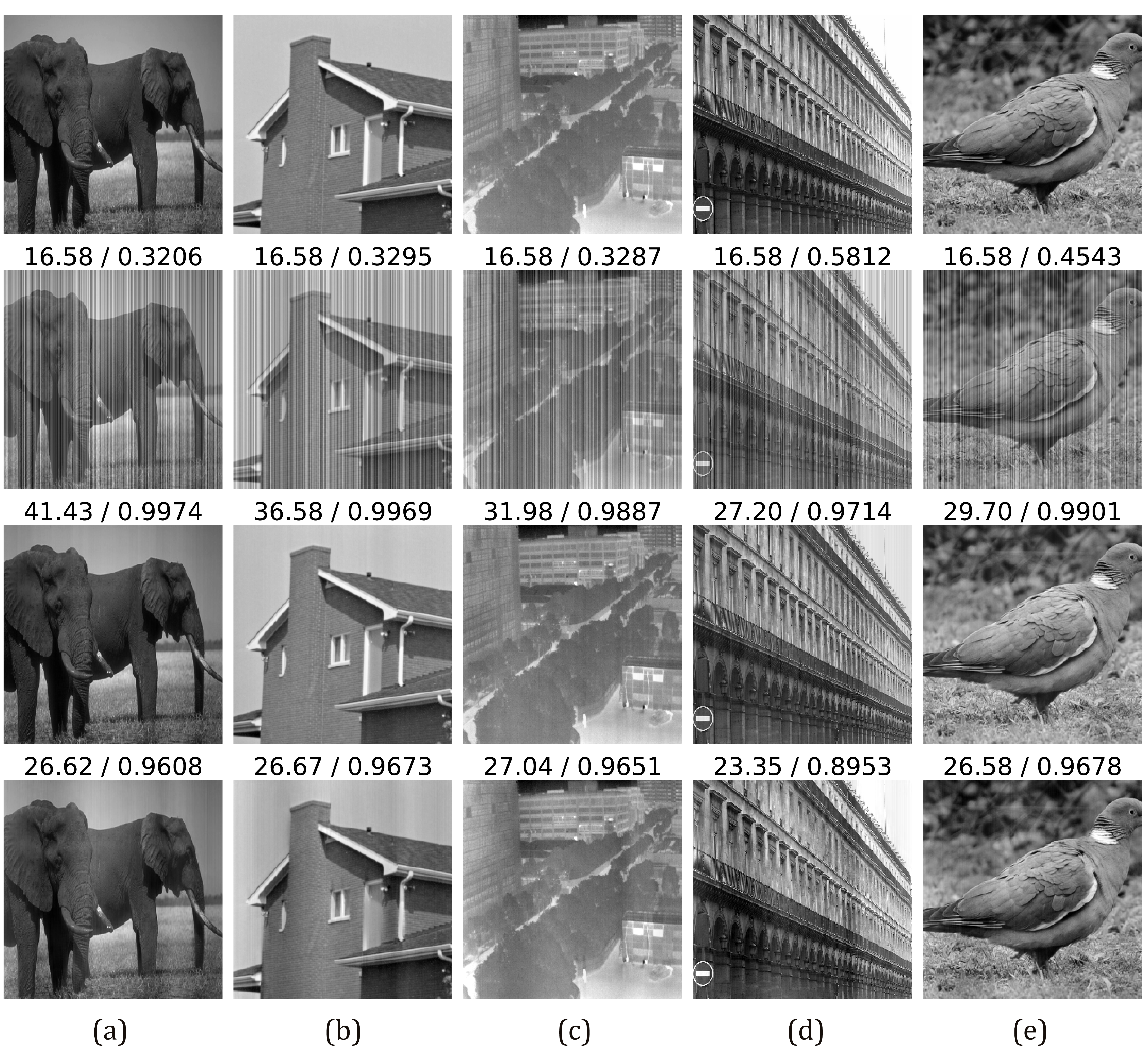}
     \caption{Destriping results comparison on a test image from each dataset with noise level 0.15. The numbers above the image indicate PSNR/SSIM values. From top to bottom: original ground truth image, degraded image, \ac{dinr} result, \ac{snrwdnn} result. (a) BSDS100; (b) Set12; (c) INFRARED100; (d) Urban100; (e) Linnaeus 5.}
\label{fig:cleanDatasets}
 \end{figure*}
 
 \begin{figure*}[htb]
     \centering
     \includegraphics[width=\textwidth]{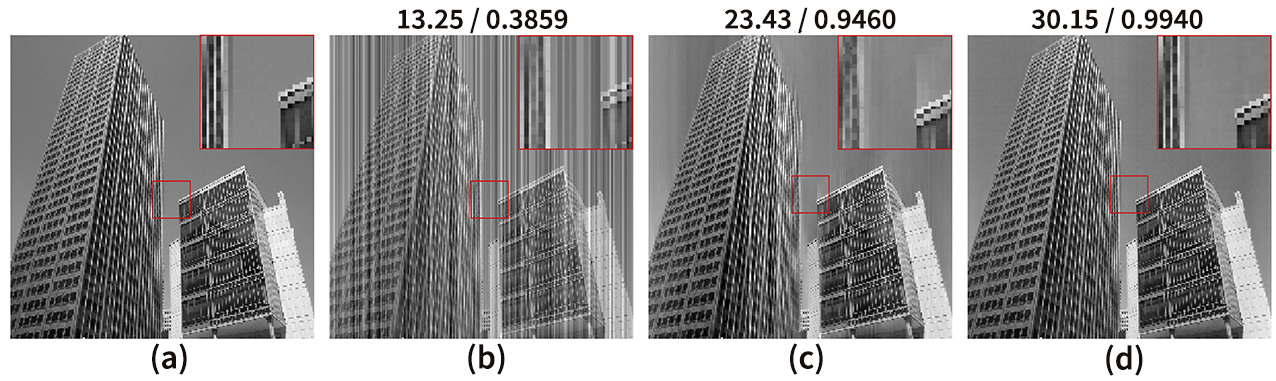}
     \label{fig:my_label}
 \end{figure*}


 \begin{figure*}[htb]
     \centering
     \vspace{-0.4cm}
     \includegraphics[width =0.82\textwidth]{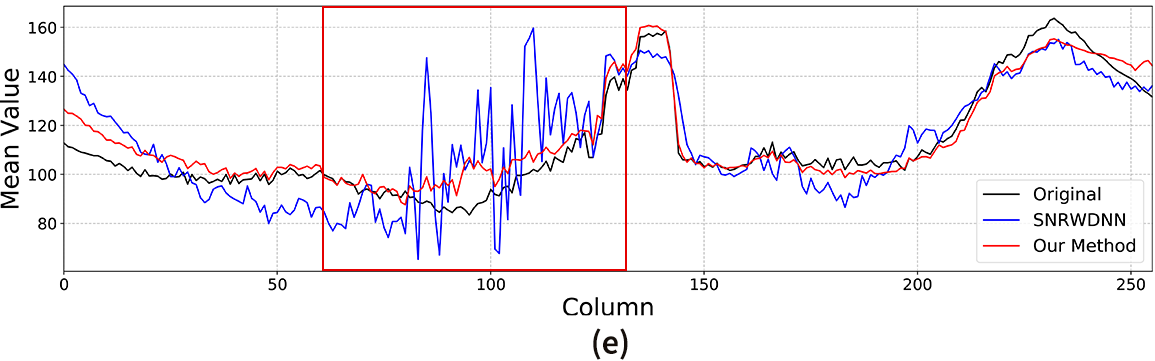}
     \label{fig:compImageGraph}
 \end{figure*}
 
 \begin{figure*}[htb]
 \vspace{-0.4cm}
     \centering
     \includegraphics[width=\textwidth]{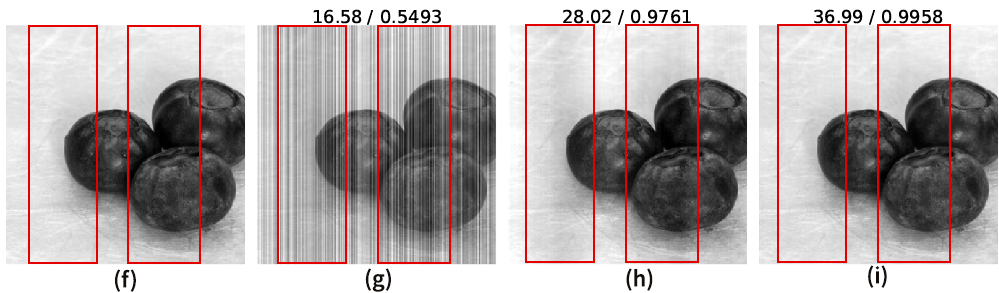}
     \label{fig:berry_cropped}
 \end{figure*}


 \begin{figure*}[!htb]
     \centering
     \vspace{-0.4cm}
     \includegraphics[width =0.82\textwidth]{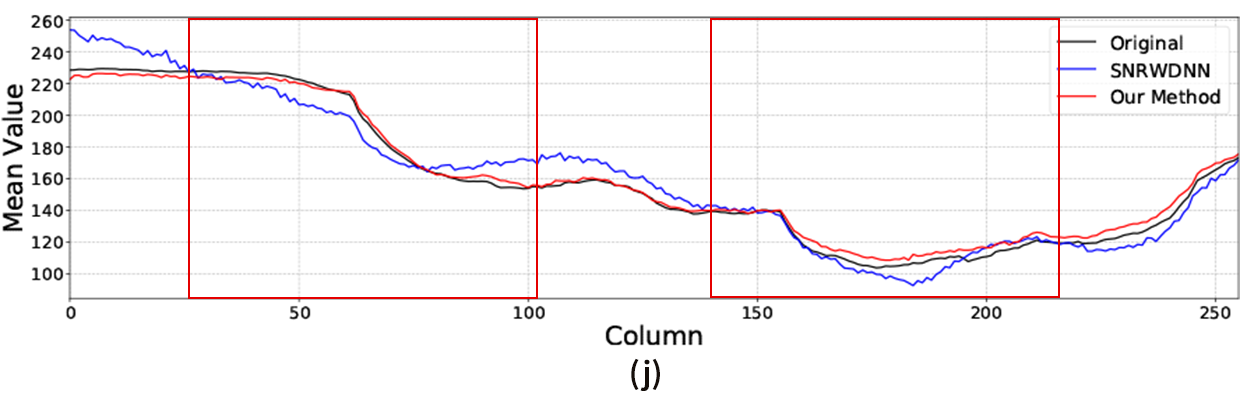}
     \caption{Destriping results comparison against SNRWDNN (a) ground truth Urban100 image; (b) image degraded with noise level 0.22; (c) result of SNRWDNN; (d) result of the proposed DINR; (e) column mean evaluation between the destriping results estimated by the proposed DINR, SNRWDNN, and original clean image; (f) ground truth Linnaeus 5 image; (g) image degraded with noise level 0.15; (h) result of SNRWDNN; (i) result of DINR; (j) same as (e).}
     \label{fig:berryImageGraph}
 \end{figure*}

\section{Evaluation Results}
\label{sec:results}

In this section, we illustrate the quantitative and qualitative performance evaluation of the \ac{dinr} method\footnote{The code for running the experiments is available at \url{https://github.com/ZeshanFayyaz/StripeNoise}}. We utilize the performance metrics of \ac{psnr} \cite{5596999} and \ac{ssim} \cite{wang2004image} \cite{hore2010image} to assess the destriping capabilities of \ac{dinr} in comparison to eight prior methods, some of which are currently state-of-the-art to the best of our knowledge. 
We begin this section by describing the quantitative evaluation indexes. Further, we describe the datasets used and an ablation study. We then compare \ac{dinr} to prior work.  


\subsection{Image Quality Metrics}

In all further experiments, we verify and compare the effectiveness of the proposed DINR model using quantitative evaluation metrics such as \ac{psnr} and \ac{ssim} \cite{hore2010image}. Given a ground-truth image $f$ and a degraded test image $g$, with dimensions of $M \times N$, the PSNR between $f$ and $g$ is defined by:

\begin{equation}
    \text{PSNR}(f,g) = 10\log_{10}(255^{2}/\text{MSE}(f,g))
\end{equation}

where, 

\begin{equation}
    \text{MSE}(f,g) = \frac{1}{MN}\sum_{i=1}^{M}\sum_{j=1}^{N} (f_{ij} - g_{ij})^2
\end{equation}

The PSNR value approaches infinity as MSE approaches zero. This indicates that a higher PSNR provides higher image quality. SSIM was developed by Wang et al. \cite{wang2004image} and is used to measure the similarity between two images. SSIM models image distortion as a combination of three factors that are loss of correlation, luminance distortion, and contrast distortion \cite{hore2010image}. Unlike PSNR, SSIM is based on visible structures in the image.

\subsection{Datasets and Ablation Study}

The publicly available datasets BSDS500 \cite{luo2015learning} and Linnaeus 5 \cite{chaladze2017linnaeus} are used to train the \ac{dinr} model. They total 6,300 images, which are split 85\% and 15\% for training and validation, respectively. These images are corrupted using stripe noise with noise intensity ($\beta$) from 0 to 0.25 according to \eqref{eq:noise} and \eqref{eq:betaDist} and tested against images with the same stripe noise intensity. The maximum number of training epochs is set at 100, with a batch size of 50. The training phase only takes about 1.5 hours on a single NVidia Quadro RTX 8000 GPU. 

For evaluation we used several different datasets including: Set12~\cite{zhang2017beyond}, BSDS100~\cite{martin2001database}, INFRARED100, Linnaeus 5, and Urban100~\cite{huang2015single}. We also evaluate the algorithm over a variety of noise intensities ($\beta$ = 0.05, 0.15, and 0.25).

We begin this section by presenting an ablation study to find the best number of iterations (i.e., unfolding layers) for our \ac{dinr} algorithm. The number of BiGRU layers examined was from 6 to 20. An excerpt of the results is summarized in Table \ref{tbl:ablation} based on the PSNR and SSIM evaluation on all five datasets using $\beta$ = 0.05 (low noise), 0.15 (moderate noise), and 0.25 (high noise). As the number of BiGRU layers increases above 15, the performance decreases on images with high-level intensity noise. On average, across all datasets and $\beta$ values, 15 BiGRU layers perform the best.


\subsection{Performance Comparison to Prior Work}

We start by evaluating the performance of \ac{dinr} compared to prior work over the Set12 test dataset. This dataset was used in prior work for performance evaluation and contains only 12 images. Table \ref{tbl:results12Data} depicts the mean PSNR and SSIM values for the degraded images, as well as the predicted images for each destriping method for various levels of noise intensity. We test our method on light noise ($\beta$ = 0.06) up to distinct high intensity noise ($\beta$ = 0.22). The model with the best-performance is depicted in bold. As can be seen from Table \ref{tbl:results12Data}, our proposed \ac{dinr} method outperforms all prior methods in terms of PSNR and SSIM. Moreover, as noise intensity increases, the performance gap between \ac{dinr} and the next best algorithm widens (about 16\% higher PSNR at $\beta$ = 0.22).

Since Set12 has only 12 images, we compare the performance of \ac{dinr} with the prior SOTA \ac{snrwdnn} over all 5 test datasets. These results are summarized in Table \ref{tbl:resultsAllDataset}, where the best performing method is
outlined in bold. For all test datasets, our \ac{dinr} model achieves significantly higher PSNR and SSIM compared to \ac{snrwdnn}. Moreover, we observe that this gap widens for higher intensity noise. This implies that the proposed method effectively distinguishes the noise component and preserves details during testing. As the proposed model produces significantly higher SSIM values, we conclude that the destriped result is more close to the original image in human perception.

For a qualitative performance evaluation, we examine the destriping capabilities of the proposed DINR algorithm against the previous SOTA, SNRWDNN. Fig.~\ref{fig:cleanDatasets} depicts a sample image randomly selected from each of the five datasets, with $\beta = 0.15$ noise intensity superimposed. As can be noted for all examples, DINR achieves a higher PSNR and SSIM value, indicating a closer resemblance to the ground truth. Qualitatively, DINR can be seen to preserve original intensities and details. Other algorithms tend to lose effectiveness at higher levels of noise, but DINR removes high-intensity regions of noise just as well as low noise. For all instances, DINR outperforms SNRWDNN extensively in PSNR and SSIM. The proposed method preserves complex details and does not over-smooth the predicted image, as can be seen in Fig.~\ref{fig:cleanDatasets}(b) and Fig.~\ref{fig:cleanDatasets}(c). An example can be seen by the destriping results of \ac{snrwdnn} in Fig.~\ref{fig:berryImageGraph}(c), which displays gray bands in the destriped image, where the high-intensity noise was.

A sample image found in Urban100 and Linnaeus 5 is shown in Fig.~\ref{fig:berryImageGraph}. Demonstrably, there are differences in the destriping results of \ac{dinr} and \ac{snrwdnn}. Specifically, there are visible residual noise artifacts in the image cleaned by \ac{snrwdnn}, while \ac{dinr} has very few artifacts. Fig.~\ref{fig:berryImageGraph}(e) illustrates the column-by-column mean pixel value. A predicted image may be considered denoised based on how closely it follows the original curve. It can be seen from columns 80-125 that the \ac{snrwdnn} model fails to track the original curve. This can be observed in Fig.~\ref{fig:berryImageGraph}(c), and we highlight the columns where \ac{snrwdnn} demonstrates a lack of detail preservation.

Similarly, the column-by-column mean pixel value of the Linnaeus 5 test image can be found in Fig.~\ref{fig:berryImageGraph}(j). The destriped result of DINR closely tracks the ground truth image, depicting that the algorithm preserves intensities and details. It is useful to note the gray bands in Fig.~\ref{fig:berryImageGraph}(h), which coincide with the high-intensity stripe noise regions in Fig.~\ref{fig:berryImageGraph}(g). These inconsistent intensity changes can be found in Fig.~\ref{fig:berryImageGraph}(j) between columns 40 and 100 as well as 140 to 215. 

\section{conclusion}
\label{sec:conclusion}
We propose \ac{dinr}, a novel stripe noise removal algorithm.  Unlike existing destriping methods, \ac{dinr} utilizes deep-unfolding to iteratively destripe the noisy image column by column. During each iteration, BiGRUs are used to estimate the column noise, using information from adjacent columns to help distinguish between noise and actual pixel values. The noise estimate is improved with each iteration (i.e., layer) of the BiGRU up until the 15th layer.  Experimental results indicate that our model performs exceptionally well at high noise intensity and outperforms classical methods as well as SOTA deep learning-based methods. These outstanding results may be seen qualitatively in the preservation of complex background details, or quantitatively, as evaluated by PSNR and SSIM. 


Future work consists of making improvements to the computational complexity, as well as utilizing our iterative BiGRU deep-unfolding approach to perform excellently in other degraded image restoration tasks consisting of vertical-patterned noise and artifacts, such as rain-removal.


%


\bibliographystyle{IEEEtran}
\bibliography{refs}

\end{document}